\documentstyle[twocolumn,aps]{revtex}

\input psfig.tex

\newcommand{\ltrsim}{\mathrel{\lower .3ex \rlap{$\sim$}
\raise .5ex\hbox{$<$}}}
\newcommand{\gttrsim}{\mathrel{\lower .3ex \rlap{$\sim$}
\raise .5ex\hbox{$>$}}}

\begin{document}
\draft

\twocolumn[\hsize\textwidth\columnwidth\hsize\csname %
@twocolumnfalse\endcsname

\title{
Limits on Phase Separation for Two-Dimensional Strongly Correlated
Electrons
}

\author{
W. O. Putikka$^{a,b}$ and M. U. Luchini$^c$
}

\address{
$^a$Theoretische Physik, ETH-H\"onggerberg, CH-8093 Z\"urich, Switzerland\\
$^b$Department of Physics, The Ohio State University, Mansfield, OH 44906$^*$\\
$^c$Department of Mathematics, Imperial College, London SW7 2BZ, United
Kingdom\\
}

\maketitle
\begin{abstract}
From calculations of the high temperature series for the free energy
of the two-dimensional $t$-$J$ model we constuct series for ratios
of the free energy per hole.  The ratios can be extrapolated very
accurately  to low temperatures and used to investigate phase
separation.  Our results confirm that phase separation occurs only
for $J/t\gtrsim1.2$.  Also, the phase transition into the phase
separated state has $T_c\approx0.25J$ for large $J/t$.
\vspace{0.3in}
\end{abstract}
]

The Hubbard and $t$-$J$ models, though widely used to investigate high
temperature superconductors, 
remain controversial when doped away from
one electron per site.  The possibility that doped holes do not form a
uniform phase but instead phase separate into distinct high
and low density regions on the lattice is an important issue that has
proved difficult to settle\cite{emery,putikka,kohno,hellberg,shih,calandra,cosentini}.  
Phase separation for physical choices of
model parameters would imply more complicated models of 2D strongly 
correlated electrons are required to describe high temperature
superconductors.  Stability of a uniform density phase would leave open
the possibility that simple models contain the relevant physics without
additional terms.

While experiments have clearly observed phase separation in a few
high-$T_c$ systems, notably oxygen overdoped La$_2$CuO$_{4+\delta}$
with mobile interstitial oxygen atoms\cite{jorgensen}, 
phase separation does not seem
to be a universal feature of the cuprates.  However, the mechanism of
phase separation causes holes to feel a net attraction, a possible 
precursor for the formation of stripe phases or superconductivity.
Finding an attractive interaction for holes in models that have
predominantly strong repulsive interactions is not easy, and all known
possibilities deserve thorough investigation.

To investigate the properties of phase separation
we have calculated the high temperature series for the 2D $t$-$J$ model free
energy to 10th order in inverse temperature.
The Hamiltonian for the
$t$-$J$ model is
\begin{equation}
H=-t\sum_{\langle ij\rangle,\sigma}\left(c_{i\sigma}^{\dagger}
c_{j\sigma} + c_{j\sigma}^{\dagger}c_{i\sigma}\right)+J\sum_{
\langle ij\rangle}\left({\bf S}_i\cdot{\bf S}_j-{1\over4}n_in_j\right),
\end{equation}
where the sums are over pairs of nearest neighbor sites and
the Hilbert space is restricted to states with no doubly occupied sites.  
The series is generated for a 2D square lattice.

To determine the stability of the uniform phase we would like to
investigate the ground state energy per hole given by
\begin{equation}
e(\delta) = {E_0(\delta)-E_0^{AF}\over\delta},
\end{equation}
introduced by Emery, Kivelson and Lin\cite{emery}.  
Here $E_0(\delta)$ is the ground state energy
per site of the uniform phase for hole doping $\delta$ and
$E_0^{AF}=-1.16944J$ is the ground state energy per site for the
Heisenberg model\cite{calandra,sandvik} 
where $\delta=0$.  If $e(\delta)$ is a monotonically
increasing function of $\delta$ the uniform phase is stable.  If
$e(\delta)$ is constant or decreasing for a range of dopings the uniform
phase is unstable for those values of $\delta$.

There are two main difficulties encountered in calculating $e(\delta)$
from numerical measurements (exact diagonalization, quantum Monte Carlo
or Green's function Monte Carlo)
of $E_0(\delta)$.  The first is that
$e(\delta)$ requires the subtraction of two large numbers 
$E_0(\delta)$ and $E_0^{AF}$ to determine
a small number which is then divided by $\delta$, another small
number.  Given statistical uncertainty in numerically
determining $E_0(\delta)$
($E_0^{AF}$ is essentially exact in comparison) this is
a difficult task, especially for $\delta\ll 1$.  The second difficulty
is that numerical calculations are done on small clusters.
Systematic errors in $E_0(\delta)$ are tough to estimate without
knowing the finite size scaling of the data and whether the cluster
sizes considered are large enough to be in the scaling 
limit\cite{calandra,cosentini}.  
In addition to these
difficulties, phase separation is favored on small clusters for
$\delta\ll 1$.  The reduction in ground state energy due to the
kinetic energy of the holes, which disfavors phase separation, is not as large on
a small cluster as it is for an infinite lattice.  On a small cluster
the electron system reduces its energy more through local interactions,
which for the $t$-$J$ model are attractive interactions for antiparallel
spins due to the $J$ term in the Hamiltonian.

High temperature series provide a means to avoid these difficulties.  We
generalize $e(\delta)$ to $T>0$ by
\begin{equation}
f(\delta,T) = {F(\delta,T)-F^{AF}(T)\over\delta},
\end{equation}
where we have replaced the ground state energy per site by the
free energy per site and $\lim_{T\rightarrow 0}
f(\delta,T)=e(\delta)$.  This replaces the difficulties mentioned above
by the need to analytically continue the series to low temperatures.
For $J/t\ltrsim 1$ and $\delta\ll 1$ we find ratios $f(\delta_2)/f(
\delta_1)$ for two closely spaced dopings $\delta_1$ and $\delta_2$
are the best quantities to extrapolate.  Series for ratios can be
calculated exactly from the series for $F$, avoiding the need to subtract
two large approximate numbers.  The series coefficients are also exact
for an infinite lattice so we have no explicit finite size effects.  The
ratios are extrapolated using standard Pad\'e approximants, but only {\it after}
the exact series for a given ratio is calculated.  
The doping
spacing we use is $\Delta\delta = 0.025$.

By extrapolating $f(\delta_2)/f(\delta_1)$ to $T=0$ we obtain estimates for
$e(\delta_2)/e(\delta_1)$ in the uniform phase.  Since high temperature
series start at infinite temperature and only have information for the
phase above a nonzero $T_c$, all of our results are for the uniform
phase.  A description of what happens if we try to extrapolate below
$T_c>0$ is given below.
Results for a range of dopings and $J/t$
values are shown in Fig. 1.  For the parameters considered here
$e(\delta)<0$ so that if $\delta_2>\delta_1$ and the system phase
separates we should find $e(\delta_2)/e(\delta_1)>1$ if $T_c>0$ or
$e(\delta_2)/e(\delta_1)=1$ if $T_c=0$.
If the uniform phase is stable we have
$e(\delta_2)/e(\delta_1)<1$.  The 2D $t$-$J$ model phase separates into
a phase with $\delta=0$ and a doped phase with $\delta=\delta_0$.  For
phase separation we therefore expect $e(\delta)/e(0.01)\ge 1$ immediately
upon doping.  
\begin{figure}[htb]
\centerline{\psfig{figure=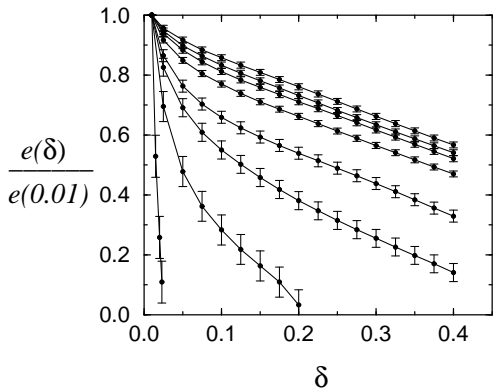,height=2in,angle=-90}}
\vspace{0.1in}
\caption{Doping dependence of the energy per hole normalized to the
energy per hole at $\delta=0.01$, plotted as a function of $\delta$ 
for different values of $J/t$.  The $J/t$ values from top to bottom 
are $J/t=0.0$, 0.2, 0.4, 0.6, 0.8, 1.0, 1.1 and 1.2.  The monotonic 
decrease of the ratio with increasing $\delta$ indicates the uniform 
phase is stable for $J/t\ltrsim1.2$.}
\end{figure}
In Fig. 1, $e(\delta)/e(0.01)<1$ and
falls monotonically with increasing $\delta$ for all $J/t$ shown, 
indicating no instability
towards phase separation in the 2D $t$-$J$ model for $J/t\ltrsim 1.2$.

In Ref. \cite{emery}  a variational argument is used to 
support the presence of
phase separation for $J/t\ll 1$.  A variational phase separated state
was constructed from two pieces occupying different parts of the lattice:
a Heisenberg antiferromagnet for the $\delta=0$ phase and a gas of
spinless fermions for the $\delta=\delta_0$ phase.  The energy of this
state is then minimized with respect to $\delta$, giving
$E_0(\delta) = E_0^{AF}-4t\delta(1-\sqrt{B\pi J/t})$ for the phase separated
state and $\delta_0=\sqrt{BJ/\pi t}$, where $B=1.16944/2=0.58472$.  This
energy was then compared to ground state energy estimates for the
uniform phase found by considering a single hole in an antiferromagnet.
The energy for the phase separated state was found to lie below the
unform state energy for small enough $J/t$, and since the variational
energy lies above the true ground state energy the conclusion of 
Ref. \cite{emery}  was that the phase separated state 
is stable.  Extrapolating the result for a single hole to a finite density
of holes assumes the energy bands remain rigid, a feature not obvious for
a strongly correlated system.  
\begin{figure}[htb]
\centerline{\psfig{figure=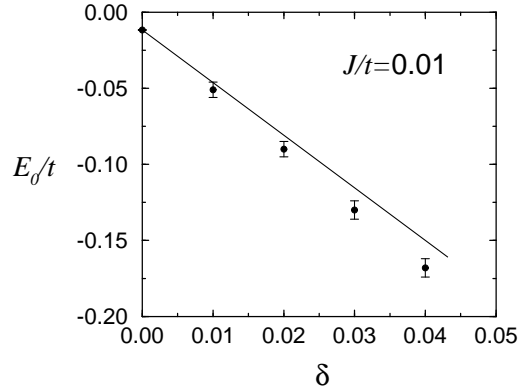,height=2in,angle=-90}}
\vspace{0.1in}
\caption{Comparison of ground state energy estimates at $J/t=0.01$.  The
solid line is a variational phase separated state 
which
extends to $\delta_0=0.04314$.  The data points are ground state energy
estimates for a uniform state calculated by extrapolating the high
temperature series for the free energy.  
At $J/t=0.01$ the variational estimate
lies above the uniform state estimate and is thus not sufficient to
show phase separation.}
\end{figure}
In Fig. 2 we compare our estimates for the
uniform ground state energy to the phase separated variational ground
state energy at $J/t=0.01$.  We find that our energies lie below the 
variational energy for $\delta<\delta_0$.  Note that from this result we
cannot conclude that the uniform state is stable, but only that the
variational state discussed in Ref. \cite{emery} 
is not sufficient to show phase
separation at $J/t\ll 1$.
\begin{figure}[htb]
\centerline{\psfig{figure=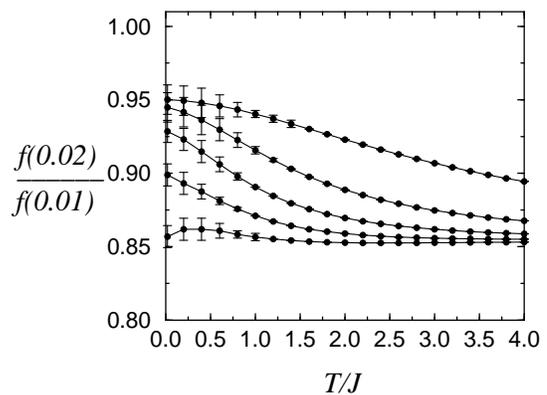,height=2in,angle=-90}}
\vspace{0.1in}
\caption{Temperature dependence of $f(0.02)/f(0.01)$ for a range of
$J/t$ values.  From top to bottom $J/t=0.2$, 0.4, 0.6, 0.8 and 1.0.
The overall temperature dependence is fairly small, making the
temperature extrapolations more reliable.}
\end{figure}

In Fig. 3 we show the temperature dependence of $f(0.02)/f(0.01)$ for a
range of $J/t$ values.  Estimating the low $T$ behavior of this function
is our only approximation.  The weak temperature dependence for the
ratio leads us to believe our results are reliable.  
The general trends
of the data shown in Figs. 1 and 3 are due to the minimum in $E_0(\delta)$
moving to smaller $\delta$ as $J/t$ is increased, causing $e(\delta)$ to
decrease in magnitude faster than $e(0.01)$, though for the parameters
shown $e(\delta)$ and $e(0.01)$ remain negative.  For values of $J/t$
larger than shown in Figs. 1 and 3 the ratio $f(\delta_2)/f(\delta_1)$
develops a spurious pole due to the crossing of $F^{AF}$ and $F(\delta_1)$
at $T>0$.  This pole greatly degrades the accuracy of extrapolations of
the ratios at lower temperatures.  To investigate larger $J/t$ we need
another method.

The chemical potential $\mu=-\partial F/\partial\delta$ provides
another means to investigate phase separation.  
We typically find $\mu$ is more
difficult to extrapolate than $f(\delta_2)/f(\delta_1)$, with the error
in the extrapolations for $\mu$ considerably larger than for the ratio.
For $J/t\gtrsim 1.2$ we do see $\mu(\delta)$ becoming quite flat for
$\delta\ll 1$, as expected for a first order phase transition into a
phase separated state.  As the temperature is lowered, $\mu$ near
the critical point (critical doping $\delta_c$ and temperature $T_c$)
becomes flat, giving a diverging compressibility $\kappa$ at the
critical point.
Results for $\mu$
are shown in Fig. 4.  The flat region found in $\mu(\delta)$ can be used
to estimate the boundary for phase separation.  However, for larger $\delta$
distinguishing where the flat region ends is difficult, leading to
errors in the position of the phase separation boundary\cite{putikka,hellberg}.
\begin{figure}[htb]
\centerline{\psfig{figure=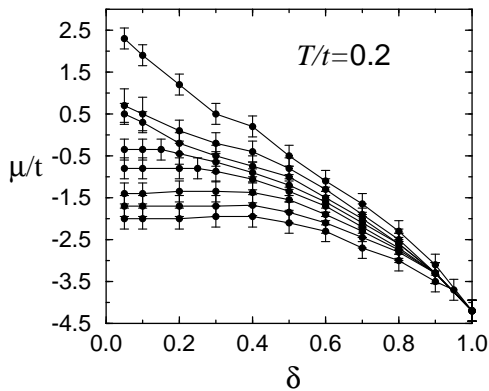,height=2in,angle=-90}}
\vspace{0.1in}
\caption{Chemical potential at $T=0.2t$ as a function of doping for a
range of $J/t$ values.  From top to bottom $J/t=0.4$, 1.0, 1.2, 1.3,
1.4, 1.6, 1.8 and 2.0.  For each $J/t\ge1.3$ a range of dopings
exists where the chemical potential is approximately constant, as 
would be expected near a critical point.}
\end{figure}

Further evidence of phase separation at large $J/t$ can be found by 
directly extrapolating $F(\delta,T)$ to estimate $E_0(\delta)$.  Fig. 5
shows results for $J/t=2.0$.  
The characteristic signature of phase
separation is the reversed curvature observed from $\delta=0$ to
$\delta\approx 0.45$.  The reversed curvature of $E_0(\delta)$ (giving
an unphysical negative compressibility) results from extrapolating
the high temperature uniform phase $F(\delta,T)$ through the $T_c>0$
phase transition for phase separation.  If $T_c=0$ we would find instead
that $E_0(\delta)$ became linear in $\delta$ in the phase separated
region.

The reversed curvature shown in Fig. 5 indicates $T_c>0$, but $T_c$ is
probably quite low.  An indirect estimate of $T_c$ can be made at large
$J/t$, above $J/t=3.4367$ where the 2D $t$-$J$ model phase separates at all
densities\cite{hellberg2}
into regions with $\delta=0$ and $\delta=1$.  Here we know
$E_0(\delta)$ for all $\delta$, since $E_0(\delta)$ is the linear
interpolation between $E_0(0)=-1.16944J$ and $E_0(1)=0$.  
\begin{figure}[htb]
\centerline{\psfig{figure=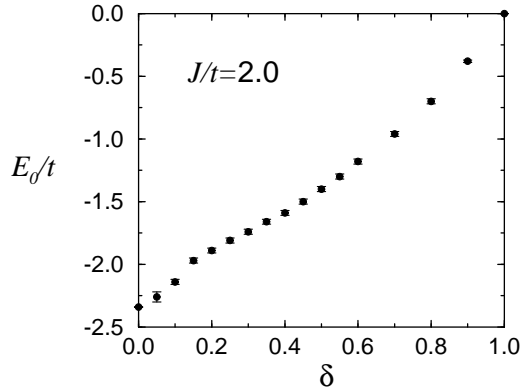,height=2in,angle=-90}}
\vspace{0.1in}
\caption{Ground state energy estimated by high temperature series as a
function of doping at $J/t=2.0$.  The reverse curvature for
$\delta\ltrsim0.45$ is due to extrapolating the high temperature free
energy through $T_c>0$ into the low temperature phase separated state.
The reverse curvature gives a negative compressibility, indicating
the uniform phase is unstable towards phase separation.}
\end{figure}
The ground
state chemical potential in this parameter range is the constant slope
of $E_0(\delta)$ with the value $\mu/t=-1.16944J/t$.  The chemical
potential hits the bottom of the tight binding band at $J/t=3.4367$
and as $J/t$ is further reduced the gain in kinetic energy
eventually limits the phase separated state to $J/t\gtrsim1.2$.
In Fig. 6 we
compare $E_0$ to $F(T)$ in the 
limit $J/t\rightarrow\infty$ with $\delta=0.5$.  
\begin{figure}[htb]
\centerline{\psfig{figure=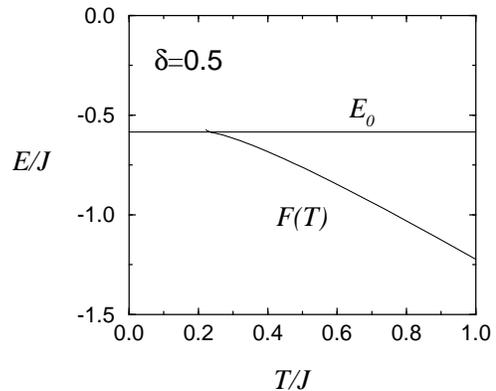,height=2in,angle=-90}}
\vspace{0.1in}
\caption{Comparison of the free energy $F(T)$ as a function of
  temperature to the known ground state energy $E_0$ for $\delta=0.5$,
$J/t\rightarrow\infty$.  The crossing of these two curves we interpret
as a phase transition at $T_c\approx0.25J$ into a phase separated state.}
\end{figure}
Comparing
the extrapolated $F(T)$ to $E_0$ we see they tend to cross at 
$T\approx0.25J$.  Since $F(T)$ must be less than $E_0$ and 
a monotonic function of $T$
this crossing cannot occur.  We interpret the tendency to
cross as a phase transition to phase separation with $T_c\approx0.25J$.

Calculations for the 2D $t$-$J$ model currently give a wide range of
minimum $J/t$ values for the presence of phase separation.  
Minimum $J/t$ values
reported in the literature are 0\cite{emery,hellberg}, 
0.5--0.6\cite{kohno,shih,calandra}, and our result of 1.2\cite{putikka}.
The latter results are in qualitative agreement in that there is a 
minimum $J/t>0$ for phase separation.  The reasons for these differences
are not clear at present.  However, while statistical errors are well under
control, systematic errors in ground state energy calculations due to
small cluster sizes are much more difficult to control.  Calculations
investigating phase separation in the 2D Hubbard model\cite{cosentini} 
find $e(\delta)$
equal to a constant for a range of dopings near half filling for the
$U=0$ tight binding model.  This spurious indication of phase separation
is due to finite size effects and is reduced for larger clusters.
Resolving the different reported results for phase sepraration 
will probably require significantly larger cluster sizes.

In conclusion, by using an analysis of the high temperature series for the
free energy per hole $f(\delta)$ at different values of $J/t$ we find that
phase sepration in the $t$-$J$ model is limited to $J/t\gtrsim 1.2$.
In addition, we find by indirect arguments that $T_c\sim0.25J$ for the
first order phase transition into the phase separated state.  Combining
this with the demonstration that phase separation can only occur at
$T=0$ for the 2D Hubbard model on a square lattice\cite{su} supports the conjecture that
the 2D Hubbard model does not phase separate for any positive $U$.
Our results
suggest phase separation in the 2D $t$-$J$ model is a classical phase
transition similar to a lattice gas with an 
attractive interaction\cite{pathria} and
that phase separation is not important for physical choices of the $t$-$J$
model parameters.  

This work was supported in part by a faculty travel grant from the
Office of International Studies at The Ohio State University (WOP),
the Swiss National Science Foundation (WOP) and by EPSRC Grant
No. GR/L86852 (MUL).  WOP thanks the ETH-Z\"urich for hospitality while
part of this work was being completed.\bigskip

\noindent
$^*$Permanent address.

\end{document}